\documentclass{article}

\usepackage{microtype}
\usepackage{graphicx}
\usepackage{booktabs}
\usepackage{array}
\usepackage{hyperref}

\usepackage[accepted]{icml2025}

\usepackage{amsmath}
\usepackage{amssymb}
\usepackage{mathtools}
\usepackage{amsthm}
\usepackage{xcolor}
\usepackage{multirow}
\usepackage{enumitem}
\usepackage{xspace}
\usepackage[capitalize,noabbrev]{cleveref}
\newif\ifhaveglossaries
\IfFileExists{glossaries.sty}{\haveglossariestrue}{\haveglossariesfalse}
\ifhaveglossaries
  \usepackage[acronym]{glossaries}%
  \setacronymstyle{long-short}%
  \glsdisablehyper
  \newcommand{\newterm}[3]{\newglossaryentry{#1}{name={#2},description={#3}}}%
\else
  \newcommand{\newacronym}[3]{\expandafter\def\csname glsshort@#1\endcsname{#2}}%
  \newcommand{\newterm}[3]{\expandafter\def\csname glsshort@#1\endcsname{#2}}%
  \newcommand{\gls}[1]{\csname glsshort@#1\endcsname}%
  \newcommand{\Gls}[1]{\csname glsshort@#1\endcsname}%
  \newcommand{\glspl}[1]{\csname glsshort@#1\endcsname s}%
\fi

\newacronym{ae}{AE}{Alfv\'en eigenmode}
\newacronym{elm}{ELM}{edge-localized mode}
\newacronym{rwm}{RWM}{resistive wall mode}
\newacronym{ntm}{NTM}{neoclassical tearing mode}
\newacronym{tm}{TM}{tearing mode}
\newacronym{mhd}{MHD}{magnetohydrodynamic}
\newacronym{ece}{ECE}{electron cyclotron emission}
\newacronym{bes}{BES}{beam emission spectroscopy}
\newacronym{qh}{QH}{quiescent H-mode}
\newacronym{doe}{DOE}{Department of Energy}
\newacronym{wpqh}{WPQH}{wide-pedestal quiescent H-mode}
\newacronym{marfe}{MARFE}{Multifaceted Asymmetric Radiation From the Edge}
\newacronym{d3d}{DIII-D}{DIII-D National Fusion Facility}

\newterm{sawteeth}{sawtooth}{sawtooth oscillations}
\newcommand{\ddd}{DIII-D\xspace}

\makeatletter
\renewcommand{\ICML@appearing}{\textit{Preprint.} Work in progress; this version
reports the catalog of \catalogdate{}. Copyright 2026 by the author(s).}
\makeatother

\theoremstyle{plain}

\theoremstyle{definition}

\theoremstyle{remark}

\newif\ifdraft
\draftfalse  
\ifdraft
  \newcommand{\note}[1]{\textcolor{blue}{\small[\textbf{Note:} #1]}}
  \newcommand{\pending}[1]{\textcolor{orange!80!black}{\small[\textbf{Pending:} #1]}}
  \newcommand{\citeph}[1]{\textcolor{blue}{\small[cite: #1]}}
\else
  \newcommand{\note}[1]{}
  \newcommand{\pending}[1]{}
  \newcommand{\citeph}[1]{}
\fi

\newif\ifhighlight
\highlightfalse  
\ifhighlight

  \DeclareRobustCommand{\flag}[1]{\textcolor{red}{\small[\textbf{Check:} #1]}}
  \renewcommand{\citeph}[1]{\textcolor{red}{\small[cite: #1]}}
\else

  \DeclareRobustCommand{\flag}[1]{}
\fi

\newcommand{\catalogdate}{October 2026}

\newcommand{\tokeye}{TokEye}
\newcommand{\seldnet}{\mbox{\texttt{ae-ours}}}  
\newcommand{\lstmnet}{\mbox{\texttt{ae-lstm}}}     
\newcommand{\rcnnet}{\mbox{\texttt{ae-rcn}}}       
\newcommand{\besnet}{\mbox{\texttt{confine-cnn}}}   
\newcommand{\elmo}{\mbox{\texttt{elm-elmo}}}       
\newcommand{\rwmbrf}{\mbox{\texttt{rwm-brf}}}      
\newcommand{\rwmours}{\mbox{\texttt{rwm-ours}}}    
\newcommand{\elmours}{\mbox{\texttt{elm-ours}}}   
\newcommand{\confours}{\mbox{\texttt{confine-ours}}} 
\newcommand{\tmours}{\mbox{\texttt{tm-ours}}}     
\newcommand{\tmcnn}{\mbox{\texttt{tm-cnn}}}  
\newcommand{\tmdsm}{\mbox{\texttt{tm-dsm}}}       
\newcommand{\tmrule}{\mbox{\texttt{tm-rule}}} 
\newcommand{\sawours}{\mbox{\texttt{saw-ours}}}   
\newcommand{\sawhl}{\mbox{\texttt{saw-hl3}}}      
\newcommand{\sawder}{\mbox{\texttt{saw-derivative}}} 
\newcommand{\detachours}{\mbox{\texttt{detach-ours}}}     
\newcommand{\detachvictor}{\mbox{\texttt{detach-frames}}} 
\newcommand{\elmdsm}{\mbox{\texttt{elm-dsm-detect}}} 

\newcommand{\nPop}{4{,}872}
\newcommand{\popShots}{185601--204999}
\newcommand{\popYears}{2021--2025}
\newcommand{\nCohort}{500}
\newcommand{\nCohortTest}{50}
\newcommand{\nCohortVal}{50}
\newcommand{\nCohortTrain}{400}
\newcommand{\nStratL}{145}
\newcommand{\nStratG}{583}
\newcommand{\nStratGc}{100}
\newcommand{\nStratR}{4{,}144}
\newcommand{\nStratRc}{255}

\newcommand{\nPhen}{seven}  

\newcommand{\auditShots}{60}

\newcommand{\aeBenchShots}{19}
\newcommand{\aeBenchFrames}{3{,}800}
\newcommand{\aeRevPrev}{68}             
\newcommand{\seldRevAUROC}{0.982}
\newcommand{\seldRevFone}{0.944}\newcommand{\seldRevAUPRC}{0.992}
\newcommand{\rcnRevAUROC}{0.836}\newcommand{\rcnRevFone}{0.774}\newcommand{\rcnRevAUPRC}{0.918}
\newcommand{\lstmRevAUROC}{0.714}\newcommand{\lstmRevFone}{0.715}\newcommand{\lstmRevAUPRC}{0.842}
\newcommand{\alwaysRevFone}{0.812}

\newcommand{\nConfMerged}{448}     
\newcommand{\nConfCurated}{401}    

\newcommand{\besMacroAUROC}{0.889}\newcommand{\besMacroAUROCci}{[0.81, 0.95]}
\newcommand{\besMacroAUPRC}{0.713}
\newcommand{\besMacroFone}{0.703}\newcommand{\besMacroCI}{[0.59, 0.79]}
\newcommand{\besOneSplit}{0.684}\newcommand{\besOneSplitSD}{0.074}  
\newcommand{\nConfBESshots}{142}\newcommand{\nConfBEStests}{200}    
\newcommand{\nConfBESfetched}{444}\newcommand{\nConfBESkept}{401}   
\newcommand{\besWPFone}{0.39}\newcommand{\besWPci}{[0.00, 0.69]}\newcommand{\nConfWPtest}{7}
\newcommand{\besPaperFone}{0.94}   
\newcommand{\nConfOursShots}{416}  
\newcommand{\confOursFone}{0.910}\newcommand{\confOursCI}{[0.882, 0.936]}
\newcommand{\confOursAUROC}{0.980}\newcommand{\confOursAUPRC}{0.948}
\newcommand{\confOursPaired}{\ensuremath{+0.207}}\newcommand{\confOursPairedci}{\ensuremath{[+0.09, +0.31]}}  
\newcommand{\confOursProtoFone}{0.909}\newcommand{\confOursProtoCI}{[0.79, 0.96]}
\newcommand{\confOursProtoAUROC}{0.987}\newcommand{\confOursProtoAUPRC}{0.939}

\newcommand{\elmoSmithP}{0.997}\newcommand{\elmoSmithR}{0.986}  
\newcommand{\elmoPaperP}{0.995}\newcommand{\elmoPaperR}{0.976}
\newcommand{\elmoShots}{73}        
\newcommand{\elmoFone}{0.842}

\newcommand{\elmoAUROC}{0.918}
\newcommand{\elmoAUPRC}{0.833}

\newcommand{\elmOursFone}{0.830}

\newcommand{\elmOursBesAUROC}{0.941}\newcommand{\elmOursBesAUPRC}{0.875}\newcommand{\elmOursBesFone}{0.845}  
\newcommand{\elmPairF}{\ensuremath{+0.002}}\newcommand{\elmPairFci}{\ensuremath{[-0.052, +0.058]}}  
\newcommand{\elmPairA}{\ensuremath{+0.023}}\newcommand{\elmPairAci}{\ensuremath{[-0.014, +0.067]}}
\newcommand{\elmDsmAUROC}{0.840}\newcommand{\elmDsmAUPRC}{0.731}\newcommand{\elmDsmFone}{0.783}  
\newcommand{\pctELMcrowd}{97}      

\newcommand{\nELMreviewed}{119}
\newcommand{\nSawReviewed}{3}

\newcommand{\nTMshots}{450}
\newcommand{\nTMpopShots}{4{,}798}
\newcommand{\pctTMunc}{34.0}      
\newcommand{\tmOursAUROC}{0.999}\newcommand{\tmOursAUPRC}{0.994}\newcommand{\tmOursFone}{0.960}
\newcommand{\tmRuleAUROC}{0.998}\newcommand{\tmRuleAUPRC}{0.986}\newcommand{\tmRuleFone}{0.940}
\newcommand{\tmCnnAUROC}{0.982}\newcommand{\tmCnnAUPRC}{0.953}\newcommand{\tmCnnFone}{0.844}
\newcommand{\tmDsmAUROC}{0.963}\newcommand{\tmDsmAUPRC}{0.871}\newcommand{\tmDsmFone}{0.799}
\newcommand{\tmOursAUROCn}{0.948}\newcommand{\tmOursAUPRCn}{0.512}\newcommand{\tmOursFonen}{0.363}
\newcommand{\tmRuleAUROCn}{0.963}\newcommand{\tmRuleAUPRCn}{0.632}\newcommand{\tmRuleFonen}{0.447}
\newcommand{\tmCnnAUROCn}{0.903}\newcommand{\tmCnnAUPRCn}{0.444}\newcommand{\tmCnnFonen}{0.435}
\newcommand{\tmDsmAUROCn}{0.839}\newcommand{\tmDsmAUPRCn}{0.297}\newcommand{\tmDsmFonen}{0.356}
\newcommand{\tmSeoRec}{12/26}\newcommand{\tmSurvRec}{16/67}
\newcommand{\tmCnnPubFone}{0.611}\newcommand{\tmDsmPubFone}{0.177}

\newcommand{\nRWMonsets}{56}\newcommand{\nRWMshots}{33}
\newcommand{\nRWMnOne}{48}\newcommand{\nRWMnTwo}{6}  
\newcommand{\nRWMbinPos}{93}\newcommand{\nRWMbinNeg}{14{,}082}  
\newcommand{\rwmOursAUROC}{0.967}\newcommand{\rwmOursAUROCci}{[0.946, 0.986]}
\newcommand{\rwmOursAUPRC}{0.523}\newcommand{\rwmOursAUPRCci}{[0.403, 0.644]}
\newcommand{\rwmOursFone}{0.529}\newcommand{\rwmOursFoneci}{[0.432, 0.609]}
\newcommand{\rwmBrfAUROC}{0.711}\newcommand{\rwmBrfAUROCci}{[0.661, 0.760]}
\newcommand{\rwmBrfAUPRC}{0.014}\newcommand{\rwmBrfAUPRCci}{[0.009, 0.021]}
\newcommand{\rwmBrfFone}{0.034}\newcommand{\rwmBrfFoneci}{[0.014, 0.060]}
\newcommand{\rwmPairA}{\ensuremath{+0.257}}\newcommand{\rwmPairAci}{\ensuremath{[+0.200, +0.315]}}  
\newcommand{\rwmPairP}{\ensuremath{+0.509}}\newcommand{\rwmPairPci}{\ensuremath{[+0.392, +0.625]}}
\newcommand{\rwmPairF}{\ensuremath{+0.495}}\newcommand{\rwmPairFci}{\ensuremath{[+0.403, +0.577]}}
\newcommand{\rwmAlwaysFone}{0.013}\newcommand{\rwmAlwaysAUPRC}{0.007}  
\newcommand{\rwmNstxAUROC}{0.918}  

\newcommand{\covDetShots}{23}         
\newcommand{\nSawShots}{13{,}319}\newcommand{\nSawCorpus}{16{,}909}  
\newcommand{\nSawPosShots}{5{,}639}  
\newcommand{\sawAbsPct}{3.9}     
\newcommand{\sawBenchShots}{386}  
\newcommand{\sawOursAUROC}{0.984}\newcommand{\sawOursAUPRC}{0.974}\newcommand{\sawOursFone}{0.958}
\newcommand{\sawHlAUROC}{0.752}\newcommand{\sawHlAUPRC}{0.489}\newcommand{\sawHlFone}{0.483}
\newcommand{\sawDerAUROC}{0.870}\newcommand{\sawDerAUPRC}{0.573}\newcommand{\sawDerFone}{0.729}
\newcommand{\sawAlwaysFone}{0.411}\newcommand{\sawAlwaysAUPRC}{0.259}
\newcommand{\sawOursAUPRCci}{[0.958, 0.984]}\newcommand{\sawOursAUROCci}{[0.976, 0.991]}\newcommand{\sawOursFoneci}{[0.941, 0.970]}
\newcommand{\sawOursCrashci}{[0.872, 0.907]}  
\newcommand{\sawDerAUPRCci}{[0.500, 0.646]}\newcommand{\sawDerAUROCci}{[0.853, 0.889]}\newcommand{\sawDerFoneci}{[0.667, 0.785]}
\newcommand{\sawDerCrashci}{[0.584, 0.727]}
\newcommand{\sawAlwaysAUPRCci}{[0.218, 0.302]}\newcommand{\sawAlwaysFoneci}{[0.358, 0.464]}
\newcommand{\sawHlInputs}{the paper's offline set: a core and an edge soft X-ray chord, EFIT-axis core ECE $T_e$, line density, $I_p$, Mirnov, stored energy, and total beam and ECH power}
\newcommand{\sawHlVsAlways}{is not clearly above}   
\newcommand{\sawHlAUPRCci}{[0.403, 0.593]}\newcommand{\sawHlAUROCci}{[0.703, 0.799]}\newcommand{\sawHlFoneci}{[0.427, 0.537]}
\newcommand{\sawHlCrashci}{[0.611, 0.746]}  
\newcommand{\sawHlWinAcc}{0.715}  
\newcommand{\sawMajWinAcc}{0.522}  
\newcommand{\sawDerCrash}{0.657}\newcommand{\sawOursCrash}{0.891}\newcommand{\sawHlCrash}{0.679}  
\newcommand{\sawRevOurs}{0.871}\newcommand{\sawRevDer}{0.654}\newcommand{\sawRevHl}{0.897}  
\newcommand{\sawElmOurs}{0}\newcommand{\sawElmDer}{13}\newcommand{\sawElmHl}{5}  

\newcommand{\nDetBins}{821}
\newcommand{\detTeAUROC}{0.95}\newcommand{\detTeCI}{[0.91, 0.98]}\newcommand{\detTeNoAnchor}{0.94}  
\newcommand{\detOursF}{0.722}\newcommand{\detVictorF}{0.54}  
\newcommand{\detOursFci}{[0.55, 0.87]}\newcommand{\detMajF}{0.34}  
\newcommand{\figOneShot}{199597}

\icmltitlerunning{Tokamak-SI: Interpreting Tokamak Discharges with Labeled Catalogs}

\begin{document}

\twocolumn[
\icmltitle{Tokamak-SI: Interpreting Tokamak Discharges with Labeled Catalogs}

\icmlsetsymbol{equal}{*}

\begin{icmlauthorlist}
\icmlauthor{Nathaniel Chen}{pton}
\icmlauthor{Alvin Garcia}{pton,orise}
\icmlauthor{Ricardo Shousha}{pppl}
\icmlauthor{Jalal Butt}{pton}
\icmlauthor{Azarakhsh Jalalvand}{pton}
\icmlauthor{Egemen Kolemen}{pton,pppl}
\end{icmlauthorlist}

\icmlaffiliation{pton}{Department of Mechanical and Aerospace Engineering, Princeton University, NJ, USA}
\icmlaffiliation{pppl}{Princeton Plasma Physics Laboratory, NJ, USA}
\icmlaffiliation{orise}{Oak Ridge Institute for Science and Education, Oak Ridge, TN, USA}

\icmlcorrespondingauthor{Nathaniel Chen}{nathaniel@princeton.edu}
\icmlcorrespondingauthor{Egemen Kolemen}{ekolemen@princeton.edu}

\icmlkeywords{Machine Learning, fusion energy, tokamak, datasets, label quality, evaluation, time series, event detection}

\vskip 0.3in
]

\printAffiliationsAndNotice{}

\begin{abstract}
Finding which phenomena occur in a tokamak discharge, and when, is often a manual task. Previous efforts at labeling events have been task-specific, each with its own corpus of data. We introduce Tokamak-SI, which combines the labels and labeling methods of many earlier studies into a single scalable, time-aligned, and categorical dataset covering observations from confinement, detachment, \glspl{ae}, \glspl{elm}, sawtooth, \glspl{tm} and \glspl{rwm} in \gls{d3d}. Models based on this, which can extend the labels to new shots, match or beat earlier detectors on its labels, including for \gls{ae}, \gls{elm}, and \gls{tm} detection. Tokamak-SI is intended to enable task-specific benchmarking, physics-searchable archives, and direct identification of key events during an experiment.
\end{abstract}

\section{Introduction}
\label{sec:intro}

A tokamak discharge, or shot, can record many diagnostic measurements which scientists inspect or pass to custom algorithms to characterize the physics of interest. As a result, fusion databases are often specialized to one purpose, such as equilibrium regression \citep{nakkina_fusion_2026,shousha_machine_2023} or event prediction \citep{zamkovska_decaf_2024,seo_avoiding_2024}. This makes analysis across disciplines disjointed, and makes it hard to search for events or to evaluate algorithms. Additionally, datasets that combine several events often add their own heuristics, and different diagnostics or heuristics applied to the same shot can give different ground truths, a problem for machine-learning workflows.

Here we attempt to invert this workflow by combining established labeling methods into one large, time-aligned dataset, with automated labels guiding expert review. The dataset makes archives of thousands of shots searchable by physics, and gives foundation models for tokamak data \citep{boschi_tokamind_2026,steiner_ignite_2026} and language-model agents labeled targets \citep{steiner_ignite_2026}.

\paragraph{A lack of aligned or standardized labels.} Interpreting a shot requires labels aligned across phenomena. Tables of Alfv\'en eigenmodes \citep{jalalvandAlfvenEigenmodeClassification2021,garcia_comparison_2023}, tearing modes
\citep{seo_avoiding_2024}, \glspl{elm} and confinement were each compiled for individual studies. Two separate sources for \glspl{elm} were found during this search, one made for the purpose of characterizing \gls{elm}-y periods \citep{oshea_automatic_2022} and the other for marking highly damaging breakthrough \glspl{elm} \citep{butt_wide-window_2025}. Benchmarks such as TokaMark \citep{rousseau_tokamark_2026} mostly target physics-informed regression, with fixed diagnostics and time windows that make new sensors hard to add. Labeling tools and standard dataset schemas have also recently been proposed and are at various stages of development
\citep{michoski_data_2025,noauthor_toktaggercitationcff_nodate,sharma_developing_nodate}.

\begin{figure*}[t!]
\centering
\includegraphics[width=\textwidth]{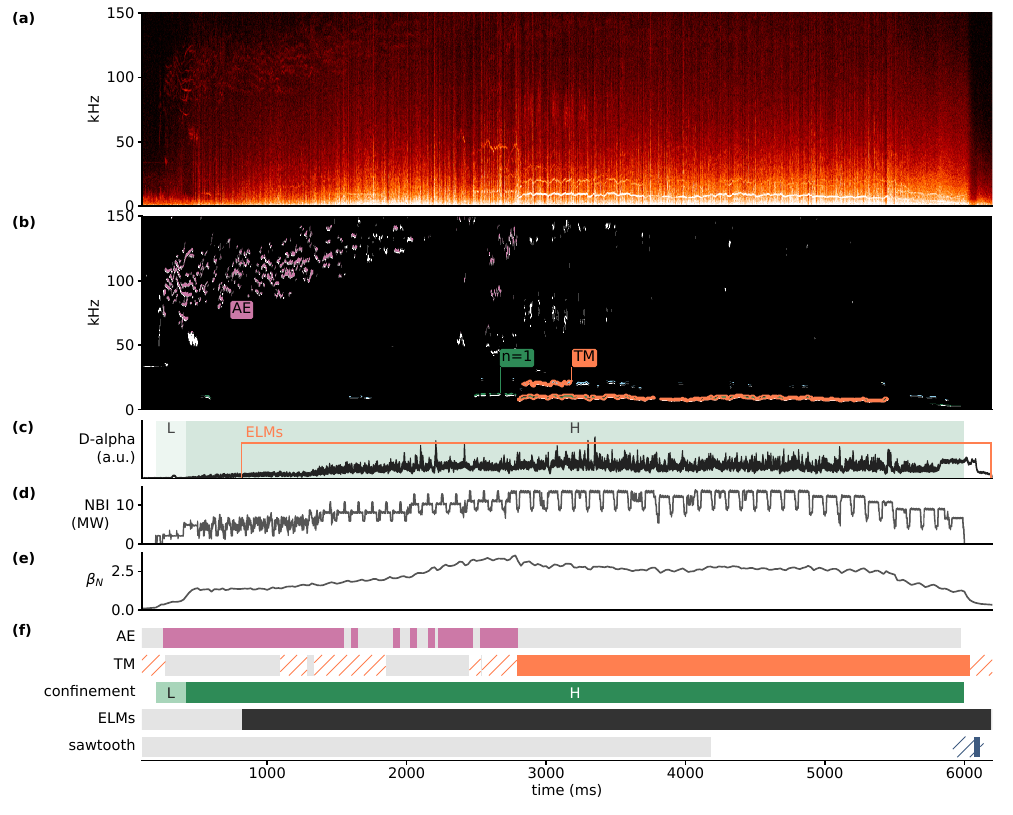}
\caption{\textbf{Shot interpretation on \gls{d3d} shot \#\figOneShot{} (0.1--6.2\,s).}
(a) Raw CO$_2$ interferometer cross-power spectrogram (chords R0$\times$V1).
(b) Automatically labeled CO$_2$ spectrogram combining \tokeye's output with toroidal mode number $n$ from the Mirnov array and labels from Tokamak-SI.
(c) D$\alpha$ with the automatically labeled \gls{elm} intervals (coral boxes) and confinement regimes.
(d) Neutral beam power.
(e) Normalized beta $\beta_N$.
(f) Automatic labels: \gls{ae}, \gls{tm}, confinement regimes, \gls{elm}, sawtooth; color present, gray absent, hatched uncertain, blank unassessed.}
\label{fig:interpreter}
\end{figure*}

\paragraph{Tokamak-SI.} Tokamak-SI, or tokamak shot interpreter, combines uniform labels for confinement regimes, detachment, \glspl{ae}, \glspl{elm}, \gls{sawteeth}, \glspl{tm} and \glspl{rwm} with a scheme for using them in analysis. Labels can be viewed on their diagnostics (\cref{fig:interpreter}), and \tokeye{} \citep{chenTokEyeFastSignal2026} is used to assign labels to coherent and transient modes in spectrograms.

\begin{figure*}[t!]
\centering
\includegraphics[width=\textwidth]{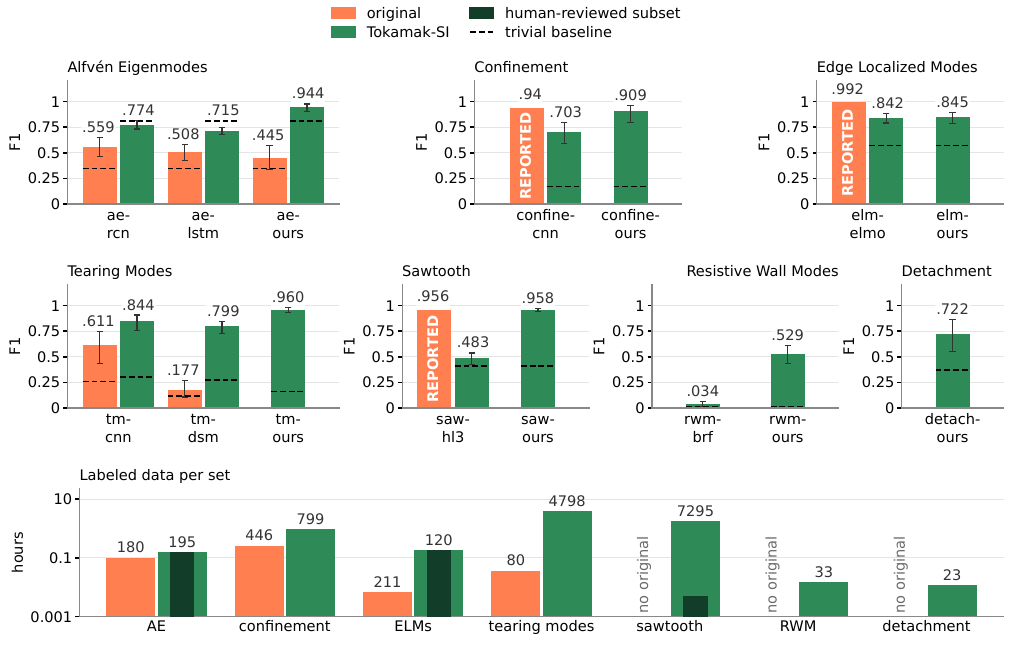}
\caption{\textbf{F1 on original and Tokamak-SI labels.} Dashes: trivial baselines. Row 3: shots with present or absent time (numbers) and hours of that time (bars).}
\label{fig:benchmarks}
\end{figure*}

\paragraph{Contributions.}
\begin{enumerate}

\item \textbf{A large, uniformly labeled dataset and framework}: aligned, categorical labels on \nPop{} discharges, with a cohort of \nCohort{} shots set aside for expert review across all phenomena.


\item \textbf{A benchmark protocol and benchmarks on prior models}: For each phenomenon, trivial baselines, detectors from earlier studies, and our models are scored on a consistent set of labels and, where original labels exist, those too (\cref{fig:benchmarks}).

\item \textbf{Automatic review of tokamak discharges}: Known activity is displayed directly on diagnostic views (\cref{fig:interpreter}).

\end{enumerate}

\section{Related Work}
\label{sec:related}

\paragraph{Large labeled datasets and label quality.} Large labeled datasets, primarily ImageNet \citep{deng_imagenet_2009} and COCO \citep{linMicrosoftCOCOCommon2015}, demonstrate that common formats for training and evaluating on large datasets can be categorical.

Subsequent work shows that labels can go beyond the classes that are present, with absent and uncertain events labeled explicitly. LVIS \citep{gupta_lvis_2019} and Open Images \citep{kuznetsova_open_2020} store verified negatives, and CheXpert's uncertain labels \citep{irvin_chexpert_2019}, PTB-XL's validation flags \citep{wagner_ptb-xl_2020} and GWTC's duty cycles \citep{collaboration_gwtc-50_2026} mark what was undecided or unobserved, as our four states and tiers do.

If an initial labeling process produces errors, especially in localization, multiple passes can be used to align datasets \citep{northcuttPervasiveLabelErrors2021,beyer_are_2020,recht_imagenet_2019}. This and AudioSet's temporally strong relabeling \citep{gemmeke_audio_2017} of its weak clip labels demonstrate recursive label improvement. Time-axis event scoring metrics additionally benefit datasets that heavily depend on forecasting or time-series classification \citep{mesaros_metrics_2016,tatbul_precision_2019}.

\paragraph{Fusion data and event labels.} Community databases usually record targeted entries, such as confinement time slices \citep{verdoolaege_updated_2021} or L--H transitions \citep{delabie_empirical_2026}. Open datasets and benchmarks (FAIR-MAST, \citealp{jackson_fair-mast_2024}; TokaMark, \citealp{rousseau_tokamark_2026}; the Fusion Equilibrium Challenge, \citealp{nakkina_fusion_2026}) usually contain established continuous physics values or simple targets. Detectors exist for \glspl{ae} \citep{jalalvandAlfvenEigenmodeClassification2021,garcia_comparison_2023}, \glspl{elm} (ELM-O, a multi-diagnostic rule; \citealp{oshea_automatic_2022}), the confinement regime (from \gls{bes}; \citealp{gillRealtimePlasmaConfinement2023}), tearing onset \citep{seoMultimodalPredictionTearing2023}, sawtooth on HL-3 \citep{ouyang2025realtime} and event chains \citep{sabbagh_disruption_2023}. We re-score most of these detectors on \gls{d3d} data with our available labels, and on their original labels where available (\cref{sec:results}). On MAST, \citet{sharma_developing_nodate} give baselines for disruption, spectrogram \gls{mhd}, confinement and \gls{elm} tasks, each on its own small label set. Spectrograms have been segmented automatically \citep{zapata-cornejo_segmentation_2024,chenTokEyeFastSignal2026}. Human-in-the-loop and software-accelerated workflows have also been proposed or provided \citep{michoski_data_2025,noauthor_toktaggercitationcff_nodate}.

\section{The Tokamak-SI Dataset}
\label{sec:catalog}

Currently, Tokamak-SI covers \nPhen{} phenomena: \glspl{ae}, \glspl{elm}, confinement, tearing modes, sawtooth, \glspl{rwm} and divertor detachment.


\subsection{Shots, tiers and alignment}
\label{sec:design}

\paragraph{Population and cohort.} The population is the \nPop{} \gls{d3d} shots numbered \popShots{} (\popYears) that pass five rules on plasma current, duration, heating and available data (\cref{app:catalog}). A cohort of \nCohort{} shots is set aside for expert review and split before any labeling into \nCohortTrain{} train, \nCohortVal{} validation and \nCohortTest{} blind test shots.

\paragraph{Provenance tiers.} Every label records its source. \emph{Gold} labels are expert-labeled data of the test split. \emph{Silver} labels are expert reviews of larger output drafts, such as an original table or a model's output. \emph{Bronze} labels are a model's output over the population, released only if the model passes a test on gold: precision and recall of at least 0.75, and a paired interval above the best trivial baseline. Labeled datasets up to bronze can be iteratively refined. While we keep original labels as tables from earlier studies, all of the standardized sets share one schema and one time axis per shot, so any two sources or phenomena compare bin by bin.

\subsection{Label sets}
\label{sec:labelsets}

Besides the task-specific categories, each label set also contains an option for uncertain observations and no observations as they can provide potentially weighted options during training. Each set also contains an option for whether the label spans a time or is an instance within a time. This covers the presence of individually identifiable events and clusters of events, which can occur in many named fusion phenomena.

\paragraph{Alfv\'en eigenmodes.} Original label sets come from \citet{jalalvandAlfvenEigenmodeClassification2021} and \citet{garcia_comparison_2023}: an expert \emph{annotation} of 180 shots, with sparse samples over 0--2\,s (\cref{app:ae}). These shots are updated to cover full discharges so that every 10\,ms frame is marked. Dense labels also cover 13 shots of the \gls{ece}-based set \citep{jalalvandAlfvenEigenmodeClassification2021} and two cohort shots, 195 shots in all.

\paragraph{Edge localized modes.} Original label sets come from \citet{oshea_automatic_2022} and a list of \gls{elm} onset times \citep{butt_wide-window_2025}. Experts reviewed \nELMreviewed{} cohort shots, starting from an automated \gls{elm} detector. Each present span is either a single \gls{elm} or a span of an \gls{elm}ing period whose individual \glspl{elm} are not separated (\cref{app:elm}).

\paragraph{Confinement.} Each time bin is L-mode, H-mode, \gls{qh} or \gls{wpqh}. Original label sets come from the confinement-mode database of \citet{gillRealtimePlasmaConfinement2023} and a second database \citep{butt_wide-window_2025} that covers \gls{wpqh} more fully.

\paragraph{Tearing modes.} Original label sets come from \citet{seo_avoiding_2024} and
the onsets used to train a tearing survival model \citep{farre-kagaInterpretingAIFusion2026}. However, these are non-specific $n{=}1$ tearing modes. So here we attempt to narrow these down as either $2/1$ or $3/2$ modes by checking poloidal phase.

\paragraph{Sawtooth.} We did not have an immediate sawtooth dataset or machine-learning dataset on hand. So we first applied a criterion based on heuristics from \gls{ece} measurements: we would classify a sawtooth if an inversion occurs on core \gls{ece} and edge \gls{ece} channels \citep{muscatello_velocity-space_2012}. It labels \nSawShots{} of the \nSawCorpus{} shots we hold. For larger sets of sawteeth, a train of at least three crashes with an inversion, 20--250\,ms apart is used as the criterion for an initial set \citep{muscatello_velocity-space_2012}.

\paragraph{Resistive wall modes.} Original label sets come from \citet{hanson_stability_2017} and \citet{hanson_variable-spectrum_2024}, which analyze onsets within discharges in 2014 and 2018.

\paragraph{Detachment.} No discrete database of detachment labels exists, so we attempt to combine methods from \citet{chen_regulation_2026}, \citet{eldon_advances_2019}, \citet{eldon_analysis_2021} and \citet{byun_radiation-front_2026} into one rule that classifies each time bin as attached, detached, or \gls{marfe} (\citealp{lipschultz_marfe_1984}; \cref{app:detach}).


\section{Benchmark Protocol and Baselines}
\label{sec:protocol}

Gold labels are not scored on all categories yet, so every score below is exploratory and no model output is yet released as bronze.

\paragraph{Baselines.} Every comparison includes an \emph{always present} baseline, which can show the effects of prevalence on prediction. Non-ML signal-processing methods are also used as simpler baselines for comparison: an \gls{elm} peak finder, which finds \glspl{elm} as D$\alpha$ bursts; \tmrule{}, the rule behind the tearing-mode labels in previous control literature \citep{seoMultimodalPredictionTearing2023,seo_avoiding_2024}; and \sawder, a derivative crash picker. Earlier papers' detectors are re-run, retrained or re-implemented (\cref{app:models}): \lstmnet{} and \rcnnet{} for \glspl{ae} \citep{garcia_comparison_2023}; \besnet{}, a 3-D CNN on \gls{bes}, for confinement \citep{gillRealtimePlasmaConfinement2023}; the rule-based \elmo{} \citep{oshea_automatic_2022} and \elmdsm{}, adapted from an \gls{elm} survival model \citep{butt_wide-window_2025}, for \glspl{elm}; the onset CNN \tmcnn{} \citep{seoMultimodalPredictionTearing2023} and the survival model \tmdsm{} \citep{farre-kagaInterpretingAIFusion2026} for tearing modes; \sawhl{} \citep{ouyang2025realtime} for sawtooth; and \rwmbrf{}, a balanced random forest \citep{piccione_predicting_2022}, for \glspl{rwm}. Our models end in \texttt{-ours} (\cref{app:models}).

\paragraph{Test sets and statistics.} \glspl{ae} are tested on the \aeBenchShots{}
\emph{benchmark shots}, those of \seldnet's \auditShots{} validation shots that no
detector trained on, using the predictions the older detectors saved.
\seldnet{} used its validation shots for early stopping and run selection, and their
\gls{ae} prevalence (0.70) differs from that of its training shots (0.41), so its
scores are optimistic. All other learned scores come from cross-validation on development shots.
Intervals are 95\% shot-bootstrap percentiles, and differences are paired.

\section{Benchmark Results}
\label{sec:results}

\Cref{tab:benchmarks} scores earlier papers' models and ours. These can also be visually compared in \cref{fig:benchmarks} (F1) and \cref{fig:benchmarks_auc} (AUROC and AUPRC).

\begin{table*}[t]
\caption{\textbf{Model scores per label set.} Bold: best per metric among
models scored on the same bins.}

\label{tab:benchmarks}
\centering
\small
\begin{tabular}{lccc}
\toprule
Model & AUROC & AUPRC & F1 \\
\midrule

\multicolumn{4}{l}{\textit{Alfv\'en eigenmodes}} \\
\seldnet{} (ours) & \textbf{\seldRevAUROC} & \textbf{\seldRevAUPRC} & \textbf{\seldRevFone} \\
\rcnnet{} \citep{garcia_comparison_2023} & \rcnRevAUROC & \rcnRevAUPRC & \rcnRevFone \\
\lstmnet{} \citep{garcia_comparison_2023} & \lstmRevAUROC & \lstmRevAUPRC & \lstmRevFone \\
\midrule

\multicolumn{4}{l}{\textit{Confinement Regimes}} \\
\besnet{} \citep{gillRealtimePlasmaConfinement2023} & \besMacroAUROC & \besMacroAUPRC & \besMacroFone \\
\besnet, as published & $\geq$0.99 & -- & \besPaperFone \\
\confours{} (ours) & \textbf{\confOursProtoAUROC} & \textbf{\confOursProtoAUPRC} & \textbf{\confOursProtoFone} \\
\midrule

\multicolumn{4}{l}{\textit{Edge Localized Modes}} \\
\elmours{} (ours) & \textbf{\elmOursBesAUROC} & \textbf{\elmOursBesAUPRC} & \textbf{\elmOursBesFone} \\
\elmo{} \citep{oshea_automatic_2022} & \elmoAUROC & \elmoAUPRC & \elmoFone \\
\elmdsm{} & \elmDsmAUROC & \elmDsmAUPRC & \elmDsmFone \\
\midrule

\multicolumn{4}{l}{\textit{Tearing Modes}} \\
\tmours{} (ours) & \textbf{\tmOursAUROC} (\tmOursAUROCn) & \textbf{\tmOursAUPRC} (\tmOursAUPRCn) & \textbf{\tmOursFone} (\tmOursFonen) \\
\tmrule{} & \tmRuleAUROC{} (\textbf{\tmRuleAUROCn}) & \tmRuleAUPRC{} (\textbf{\tmRuleAUPRCn}) & \tmRuleFone{} (\textbf{\tmRuleFonen}) \\
\tmcnn{} \citep{seoMultimodalPredictionTearing2023} & \tmCnnAUROC{} (\tmCnnAUROCn) & \tmCnnAUPRC{} (\tmCnnAUPRCn) & \tmCnnFone{} (\tmCnnFonen) \\
\tmdsm{} & \tmDsmAUROC{} (\tmDsmAUROCn) & \tmDsmAUPRC{} (\tmDsmAUPRCn) & \tmDsmFone{} (\tmDsmFonen) \\
\midrule

\multicolumn{4}{l}{\textit{Sawtooth}} \\
\sawours{} (ours) & \textbf{\sawOursAUROC} & \textbf{\sawOursAUPRC} & \textbf{\sawOursFone} \\
\sawhl{} \citep{ouyang2025realtime} & \sawHlAUROC & \sawHlAUPRC & \sawHlFone \\
\sawder{} & \sawDerAUROC & \sawDerAUPRC & \sawDerFone \\
\midrule

\multicolumn{4}{l}{\textit{Resistive Wall Modes}} \\
\rwmours{} (ours) & \textbf{\rwmOursAUROC} & \textbf{\rwmOursAUPRC} & \textbf{\rwmOursFone} \\
\rwmbrf{} \citep{piccione_predicting_2022} & \rwmBrfAUROC & \rwmBrfAUPRC & \rwmBrfFone \\
\rwmbrf, as published & \rwmNstxAUROC & -- & -- \\

\bottomrule
\end{tabular}

\end{table*}

\subsection{Alfv\'en eigenmodes}

Against the dense labels \seldnet{} leads on every metric (F1 \seldRevFone{} against
\rcnRevFone{} and \lstmRevFone; \cref{tab:ae_full}). With \aeRevPrev\% of
frames present, neither older detector beats \emph{always present} (F1
\alwaysRevFone) at its own threshold, and even thresholds tuned on the scored frames
give them F1 of 0.82--0.87.


\subsection{Confinement}

Pooled over five random by-shot splits (\nConfBESshots{} shots),
\besnet{} reaches macro-F1 \besMacroFone{} \besMacroCI, against \besPaperFone{} in its paper
on 44 test shots. No protocol factor we tested moves the score by more than 0.05, so
the gap is unexplained (\cref{tab:conf_bes}). \confours{}, a 1-D U-Net over slow
signals without \gls{bes}, reaches \confOursProtoFone{} \confOursProtoCI{} on the same
windows (paired difference \confOursPaired{} \confOursPairedci) and \confOursFone{}
\confOursCI{} on \nConfOursShots{} shots in run-day folds. It is experimental, since
its labels were set from the traces it reads.

\subsection{Edge localized modes}

On the hand-labeled windows of \citet{oshea_automatic_2022}, \elmo{} reproduces its published precision and
recall (\elmoSmithP{} and \elmoSmithR, against \elmoPaperP{} and \elmoPaperR); the
windows were selected around known \glspl{elm}, so these scores hold only on them. Our labels
pose a different task, \gls{elm}y-period detection. On the \elmoShots{} reviewed shots
with \gls{bes}, \elmours{}, a 1-D U-Net over D$\alpha$ and density without \gls{bes}, reaches
F1 \elmOursBesFone{} against \elmo's \elmoFone{} (difference \elmPairF{} \elmPairFci;
\cref{tab:elm_benchmark}), and it also runs on shots without \gls{bes} (F1 \elmOursFone{} on
all \nELMreviewed{} reviewed cohort shots; \cref{fig:elm_examples}).

\subsection{Other label sets}

\paragraph{Tearing modes.} Against their original targets, the published
forecasters \tmcnn{} and \tmdsm{} reach F1 \tmCnnPubFone{} (training overlap unknown) and \tmDsmPubFone. Retrained
to detect a mode at the current time (\cref{app:models}), they reach F1
\tmCnnFone{} and \tmDsmFone{} on our labels, on the shots with their inputs.
\tmours{} reaches F1 \tmOursFone{} and AUROC \tmOursAUROC{} on \nTMshots{}
shots (\cref{tab:tm_benchmark,fig:tm_examples}).

\paragraph{Sawtooth.} A sawtooth needs a $q{=}1$ surface, so the benchmark also scores
observable time with no $q{=}1$ surface as absent (\cref{app:saw}). Out of fold on
\sawBenchShots{} training shots, \sawours{} leads on presence F1, AUROC and crash
F1 within 2\,ms (\sawOursFone, \sawOursAUROC{} and \sawOursCrash, against
\sawDerFone, \sawDerAUROC{} and \sawDerCrash{} for \sawder; \cref{tab:saw}). \sawhl's
F1 (\sawHlFone) \sawHlVsAlways{} always present (\sawAlwaysFone), though on its
paper's three-class window task it beats the majority class (window accuracy \sawHlWinAcc{} against \sawMajWinAcc). 

\paragraph{Resistive wall modes.} Both models are balanced random forests
\citep{piccione_predicting_2022} that classify each 10\,ms bin of the listed shots, scored against mode durations
derived from the onsets (\cref{app:rwm}). \rwmbrf{} reads equilibrium scalars, the $n{=}1$ and $n{=}2$ magnetic
RMS and rotation, and reaches F1 \rwmBrfFone, against \rwmAlwaysFone{} for always
present. \rwmours{} adds the $n{=}1$ $\delta B_p$ of the midplane poloidal-field probes,
I-coil activity and rotation at $q{=}2$, and reaches F1 \rwmOursFone{} \rwmOursFoneci{} and AUROC \rwmOursAUROC{}
(\cref{tab:rwm}).

\paragraph{Detachment.} With no independent reference, we report checks. The divertor
tangential camera label ranks divertor Thomson $T_e$ with AUROC \detTeAUROC{} \detTeCI{} on
\covDetShots{} shots (\detTeNoAnchor{} without the calibration shot),
measured before the L-mode correction (\cref{app:detach},
\cref{fig:detach_timeline}). Trained on the labels with shot-grouped folds,
\detachours{} reaches macro-F1 \detOursF{} \detOursFci{} (majority class \detMajF),
which measures consistency rather than accuracy.

\section{Limitations and Roadmap}
\label{sec:limitations}

\paragraph{Limitations.}


The \gls{ae} labels
started from the annotation, use only present and absent, predate the population
and lie only on \gls{ae} shots, so false alarms on \gls{ae}-free shots and expert agreement are
unmeasured. Confinement, \gls{rwm}, tearing-mode and detachment shot are in the process of full review.

\paragraph{Roadmap.} Next come a second \gls{ae} reader, gold labels on the consistent 50 blind shots across all categories,
cohort review in random order to give population base rates, re-scoring every method
on gold, and a comparison of fully labeled against single-phenomenon training at
equal expert cost. We will release labels, splits, code and weights
(\cref{app:datasheet}).

\section*{Acknowledgements}

We thank Jeremy Hanson for datasets on the \gls{d3d} \gls{rwm}, Hiro Kaga-Farre and Kei Yasoda for guidance on \gls{tm} analysis methods, Kouroche Bouchiat, Sangkyeun Kim, Jayson Barr and Minseok Kim for their help and datasets, and Lawrence Livermore National Laboratory for the divertor tangential camera (TangTV) access. The authors gratefully acknowledge the collaboration of the \ddd Team.

This material is based upon work supported by the U.S. \gls{doe}, Office of Science, Office of Fusion Energy Sciences, using the \ddd National Fusion Facility, a \gls{doe} Office of Science user facility, under Award Nos.\ DE-FC02-04ER54698, DE-SC0014664, DE-AC02-09CH11466, DE-SC0024527, DE-SC0026429 and DE-AC52-07NA27344. Additional support was provided by the Princeton Laboratory for Artificial Intelligence under Award No.\ 2025-97.

\paragraph{Disclaimer.} This report was prepared as an account of work sponsored by an
agency of the United States Government. Neither the United States Government nor any
agency thereof, nor any of their employees, makes any warranty, express or implied, or
assumes any legal liability or responsibility for the accuracy, completeness, or
usefulness of any information, apparatus, product, or process disclosed, or represents
that its use would not infringe privately owned rights. Reference herein to any
specific commercial product, process, or service by trade name, trademark,
manufacturer, or otherwise does not necessarily constitute or imply its endorsement,
recommendation, or favoring by the United States Government or any agency thereof. The
views and opinions of authors expressed herein do not necessarily state or reflect
those of the United States Government or any agency thereof.

\section*{Impact Statement}

This work aims to make fusion experiments machine-interpretable, to speed a candidate
source of low-carbon energy. The dataset holds plasma physics data and no personal
information. Operator logbook text is used only through a lexicon, and annotator
identities will be pseudonymized in the release. The main risk is over-trust: if
people or language models read model outputs as facts, detector errors could reach
scientific conclusions. So outputs that fail their bar ship as suggestions, never as
labels; every label carries its source and coverage; and the language interface
separates observations from text mentions and forecasts (\cref{app:llm}).

\bibliography{references}
\bibliographystyle{icml2025}

\newpage
\appendix
\onecolumn
\section{Catalog Details}
\label{app:catalog}

\paragraph{Population rules.} A shot enters the population if it (1) is a plasma shot
with $|I_p|\geq0.5$\,MA, a pulse of at least 2\,s and peak beam power of at least
1\,MW or any ECH power; (2) is not machine time (no start-up or power-system-test
logbook title); (3) has magnetics, \gls{ece} and filterscope data each spanning at
least 2\,s; (4) has a flat-top of at least 1\,s inside the assessed window with
$|I_p|$ at least 0.8 of its plateau (the window ends if the current falls below 30\%
of the plateau and then rises above 60\%); and (5) is not a runaway-electron plateau
(13 shots removed).

\paragraph{Cohort.} All \nStratL{} shots linked to a paper, \nStratGc{} of the
\nStratG{} shots in an original table and \nStratRc{} of the other \nStratR, drawn at
random within each year (\popYears) and source group $h$, with weights $N_h/n_h$
(population over cohort count).

\paragraph{Schema.} Intervals are
\texttt{shot,category,t\_start,t\_end,confidence[,attrs]}, times in ms, category
0/1/2/3 for absent/present/uncertain/not observable; time with no row is not
assessed. Each row records its tier and source, and each phenomenon is also exported
on a 50\,ms grid.

\begin{table}[ht]
\caption{\textbf{Labels in Tokamak-SI.} ``Labeled'' counts the \gls{ae} dense labels
and, otherwise, every shot of the original, rule or detector source. Confinement has
no absent class, so the last two columns do not apply. \gls{rwm} onsets are points.
Detachment counts detached bins as present.}
\label{tab:datasets}
\centering
\small
\begin{tabular}{lcccc}
\toprule
Phenomenon & Labeled & Expert-reviewed & Positive & Present (s) \\
\midrule
\gls{ae} & 195 & 195 & 194 & 350.2 \\
Tearing mode & 28{,}521 & 0 & 4{,}820 & 2902.3 \\
Confinement (4 classes, merged) & \nConfMerged & 0 & -- & -- \\
\gls{elm}ing & 676 & 120 & 517 & 440.3 \\
Sawtooth & 13{,}319 & \nSawReviewed & 5{,}639 & 5229.4 \\
\gls{rwm} (onsets) & 33 & 0 & 33 & -- \\
Detachment & 23 & 0 & 12 & 18.4 \\
\bottomrule
\end{tabular}

\end{table}

\section{How Each Label Is Made}
\label{app:labels}\label{app:definitions}

\Cref{tab:definitions} summarizes each label. Original tables are converted as
published: a bin is present where a positive sample or interval overlaps it and
absent only where the source gives zeros. Reviewers mark time on fixed panels;
unmarked time is not assessed.

\begin{table}[ht]
\caption{Definition summary. ``Not observable'' is set when the deciding diagnostics
are missing.}
\label{tab:definitions}
\centering\small
\begin{tabular}{p{2.2cm}p{6.0cm}p{4.5cm}p{2.4cm}}
\toprule
Phenomenon & Present means & Review panels & Not observable when \\
\midrule
\gls{ae} & Coherent Alfv\'enic modes (TAE, RSAE, EAE, BAE, low-frequency) &
CO\textsubscript{2} R0$\times$V1/V2/V3 cross-power; else magnetics and \gls{ece}
spectrograms & neither CO\textsubscript{2} nor magnetics and \gls{ece} \\
\gls{elm}ing & \gls{elm}ing intervals; absent inside H-mode is \gls{elm}-free &
D$\alpha$ with transient activity & no filterscopes \\
Confinement & One class per bin: L, H, \gls{qh}, \gls{wpqh}, or uncertain &
D$\alpha$, line-averaged density, $\beta_N$ & no filterscopes \\
Tearing mode & A rotating $n{=}1$ or $n{=}2$ mode: magnetic RMS above 12\,G ($n{=}1$)
or 6\,G ($n{=}2$) for 50\,ms with a rotating line & $n{=}1$ and $n{=}2$ magnetic RMS
and frequency; Mirnov spectrogram and $n$ fit; radial field for locks & gaps in the
magnetic RMS \\
Sawtooth & A train of $\geq$3 core \gls{ece} crashes with inversion, 20--250\,ms
apart; absent only where valid core \gls{ece} shows no crashes & \gls{ece} profile and
channels, D$\alpha$ & \gls{ece} cut off, invalid or missing \\
\gls{rwm} & From each listed onset until the $n{=}1$ $\delta B_p$ falls back below
threshold (10--300\,ms); uncertain 20\,ms before it & $n{=}1$ $\delta B_p$, $\beta_N$,
$l_i$, $I_p$ & no probe data \\
Detachment & Attached, detached or \gls{marfe} by the height of the C-III emission
front on the divertor tangential camera (TangTV) & TangTV with EFIT, Afrac, divertor
radiation, divertor Thomson & no valid TangTV call \\
\bottomrule
\end{tabular}
\end{table}

\subsection{Alfv\'en eigenmodes}
\label{app:ae}
\begin{figure*}[t]
\centering
\includegraphics[width=\textwidth]{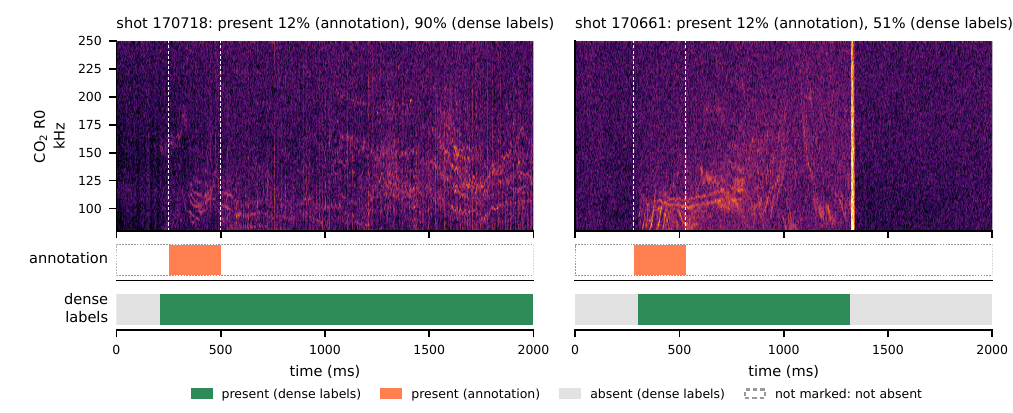}
\caption{\textbf{Annotation and dense labels} on two benchmark shots, 0--2\,s. Top:
CO\textsubscript{2} R0 spectrogram (80--250\,kHz); dashed lines bound the annotation's
marked stretch. Below: the annotation (blank: not marked) and the dense labels (gray:
absent). The annotation marks 12\% of each shot, the dense labels 90\% and 51\%. A
detector that follows the activity is scored a false positive outside the marked
stretch.}
\label{fig:ae_annotation}
\end{figure*}

The annotation covers 180 shots (170659--178879) in five \gls{ae} classes
\citep{heidbrinkBasicPhysicsAlfven} without timestamps: its samples spread evenly over 0--2\,s, and a
10\,ms frame is present when at least half its samples carry TAE, RSAE, EAE or BAE.
The dense labels mark every 10\,ms frame of 0--2\,s and, on many shots, the rest of
the shot; two experts checked them (silver). \Cref{fig:ae_annotation} shows why the
two differ.

\subsection{Edge localized modes}
\label{app:elm}
A 50\,ms bin holding an onset from \citet{butt_wide-window_2025} is present. Review
starts from a peak finder on \tokeye's transient mask of the eight D$\alpha$
filterscopes: peaks of prominence 0.03 at least 3\,ms apart, without bursts wider than
5\,ms such as gas puffs; fewer than five peaks per second is \gls{elm}-free. Experts
then marked spans in all four states (silver). As with crowds in COCO
\citep{linMicrosoftCOCOCommon2015}, a present span is one \gls{elm} or a crowd of
unseparated ones (\pctELMcrowd\% of present bins). About half the crowd edges stay
within 1\,ms of the finder's, which favors models that read D$\alpha$.

\subsection{Confinement}
\label{app:conf}
\gls{qh} (with an edge harmonic oscillation) and \gls{wpqh} (near-zero torque, wide
pedestal) are \gls{elm}-free H-modes. The label is the union of the two databases'
intervals; their one H--L conflict is uncertain, and time neither flags is not
assessed. Review starts from these intervals on the \nConfCurated{} shots that have
them and D$\alpha$, and elsewhere from a D$\alpha$ H-mode detector; \confours{} labels
the rest, unreviewed.

\subsection{Tearing modes}
\label{app:tm}
Our rule labels a mode when the $n{=}1$ magnetic RMS exceeds 12\,G
\citep{seoMultimodalPredictionTearing2023} or the $n{=}2$ RMS exceeds 6\,G for 50\,ms
with a stable rotating line of the same $n$; an $n{=}2$ mode below 0.57 of the
$n{=}1$ RMS is a harmonic and is dropped. Unsupported crossings, chirping bursts,
weaker lines and time after a lock are uncertain (\pctTMunc\% of observable time).
The archive of \citet{seo_avoiding_2024} marks a phase rather than the mode's
lifetime; within 100\,ms, the rule recovers \tmSeoRec{} of its onsets and
\tmSurvRec{} of the survival model's, starting a median 130\,ms earlier. Labels cover
\nTMshots{} development and \nTMpopShots{} population shots in 10\,ms bins.

\subsection{Sawtooth}
\label{app:saw}
A crash is a drop on adjacent core \gls{ece} channels next to a rise on outer ones
(10\,kHz, above the noise), and present is a train of at least three crashes
20--250\,ms apart \citep{muscatello_velocity-space_2012}; on 4{,}392 shots, 92.3\% of
the inversion points lie within 0.15\,m of the EFIT $q{=}1$ radius. Absent needs valid
core channels ($\rho<0.5$) with no periodic drop above 2\% over $\pm$375\,ms, which is
rare (\sawAbsPct\% of observable cohort time), so the benchmark also counts observable
time with no $q{=}1$ surface ($q_{\min}\geq1.2$ for 100\,ms) as absent. \nSawShots{}
of \nSawCorpus{} shots carry labels, \nSawPosShots{} with present time. On published
reference shots \citep{muscatello_velocity-space_2012}, a calibration step between
channels makes the rule mark the profile invalid. \nSawReviewed{} shots were reviewed
(silver).

\begin{table}[ht]
\caption{\textbf{Sawtooth benchmark}, out of fold on 2\,ms bins of the
\sawBenchShots{} training shots (three shot-grouped folds), against the rule with time
that has no $q{=}1$ surface counted absent, so it measures agreement, not accuracy.
Crash F1 counts picks within 2\,ms of a rule crash; \sawhl{} has no crash output, so
its row gates the derivative picker by its presence output. Last two columns: AUROC on
the reviewed span of 189324, and the share of 190637's reviewed \gls{elm} span called
present.}
\label{tab:saw}
\centering\small
\setlength{\tabcolsep}{3pt}
\begin{tabular}{lcccccc}
\toprule
Model & AUPRC & AUROC & Presence F1 & Crash F1 & 189324 & 190637 \\
\midrule
\sawours{} & \sawOursAUPRC{} \sawOursAUPRCci & \sawOursAUROC{} \sawOursAUROCci & \sawOursFone{} \sawOursFoneci & \sawOursCrash{} \sawOursCrashci & \sawRevOurs & \sawElmOurs\% \\
\sawhl{} & \sawHlAUPRC{} \sawHlAUPRCci & \sawHlAUROC{} \sawHlAUROCci & \sawHlFone{} \sawHlFoneci & \sawHlCrash{} \sawHlCrashci & \sawRevHl & \sawElmHl\% \\
\sawder{} & \sawDerAUPRC{} \sawDerAUPRCci & \sawDerAUROC{} \sawDerAUROCci & \sawDerFone{} \sawDerFoneci & \sawDerCrash{} \sawDerCrashci & \sawRevDer & \sawElmDer\% \\
always present & \sawAlwaysAUPRC{} \sawAlwaysAUPRCci & 0.500 & \sawAlwaysFone{} \sawAlwaysFoneci & -- & -- & 100\% \\
\bottomrule
\end{tabular}
\end{table}

\subsection{Resistive wall modes}
\label{app:rwm}
Two onset lists \citep{hanson_stability_2017,hanson_variable-spectrum_2024} hold
\nRWMonsets{} onsets on \nRWMshots{} shots from 2014 and 2018 (\nRWMnOne{} $n{=}1$ and
\nRWMnTwo{} $n{=}2$ events). Each mode's duration is measured on the $n{=}1$
$\delta B_p$ of midplane poloidal-field probes 180$^\circ$ apart: each probe difference
loses its 100\,ms running median and its I-coil and C-coil pickup, and
$2A\cos(\phi-\phi_0)$ is fitted across the probes. From an onset $o$ the mode is
present until $A$ falls back below the larger of the shot's median plus 6 MADs and
5\,G, $\beta_N$ falls by 40\%, the current ends or 300\,ms pass, for at least 10\,ms;
$n{=}2$ onsets end on the $n{=}2$ RMS. $[o{-}20\,\mathrm{ms}, o)$ and other runs above
the threshold at high $\beta_N/l_i$ are uncertain, since $\delta B_p$ alone cannot
tell an \gls{rwm} from a locked tearing mode. The rest of a listed shot's high-current
window is negative, which assumes the lists are complete.

\begin{table}[ht]
\caption{\textbf{\gls{rwm} presence} on 10\,ms bins of the listed shots, against the
durations of \cref{app:rwm}: \nRWMbinPos{} present and \nRWMbinNeg{} assumed-negative
bins, one run day held out at a time. \rwmours{} adds the $n{=}1$ $\delta B_p$, I-coil
activity and $q{=}2$ rotation to \rwmbrf's inputs; the durations end on the same
$\delta B_p$.}
\label{tab:rwm}
\centering
\small
\setlength{\tabcolsep}{4pt}
\begin{tabular}{@{}lccc@{}}
\toprule
\multicolumn{4}{@{}l}{\gls{rwm} presence, \gls{d3d} (assumed negatives)} \\
\midrule
Model & AUROC & AUPRC & F1 \\
\midrule
\shortstack[l]{\rwmours{}\\\phantom{{\scriptsize $[\,]$}}} & \shortstack{\rwmOursAUROC\\{\scriptsize \rwmOursAUROCci}} & \shortstack{\rwmOursAUPRC\\{\scriptsize \rwmOursAUPRCci}} & \shortstack{\rwmOursFone\\{\scriptsize \rwmOursFoneci}} \\
\addlinespace[2pt]
\shortstack[l]{\rwmbrf{}\\\phantom{{\scriptsize $[\,]$}}} & \shortstack{\rwmBrfAUROC\\{\scriptsize \rwmBrfAUROCci}} & \shortstack{\rwmBrfAUPRC\\{\scriptsize \rwmBrfAUPRCci}} & \shortstack{\rwmBrfFone\\{\scriptsize \rwmBrfFoneci}} \\
\addlinespace[2pt]
\shortstack[l]{\rwmours{} minus\\\rwmbrf{}} & \shortstack{\rwmPairA\\{\scriptsize \rwmPairAci}} & \shortstack{\rwmPairP\\{\scriptsize \rwmPairPci}} & \shortstack{\rwmPairF\\{\scriptsize \rwmPairFci}} \\
\addlinespace[2pt]
always present & 0.500 & \rwmAlwaysAUPRC & \rwmAlwaysFone \\
\bottomrule
\end{tabular}

\end{table}

\subsection{Detachment}
\label{app:detach}
On TangTV's C-III emission, the front height $\mathrm{DZ}=1-(Z_X-Z_E)/(Z_X-Z_S)$
places the emission front $E$ between the outer strike point $S$ and the X-point $X$:
attached below 0.35, detached from 0.5 to 1.2 and \gls{marfe} above, with cutoffs from
one H-mode shot, 201081 \citep{chen_regulation_2026}. A bin takes this call when the
strike point lies at $R\geq1.37$\,m, and is uncertain where it lies further in, where
the target probes' Afrac disagrees, or where a detached call falls in an L-mode phase.
\nDetBins{} 50\,ms bins on \covDetShots{} shots are labeled, none reviewed. All 6 of
the label's calls on published points agree
\citep{chen_regulation_2026,eldon_analysis_2021}. With five shot-grouped folds,
\detachours{} (0-D signals) and \detachvictor{} (raw TangTV frames;
\citealp{victor_identifying_2024}) reach macro-F1 \detOursF{} \detOursFci{} (majority
class \detMajF) and \detVictorF{} [0.25, 0.75] (majority class 0.45), which measures
consistency, not accuracy.

\begin{figure*}[!htbp]
\centering
\includegraphics[width=\textwidth]{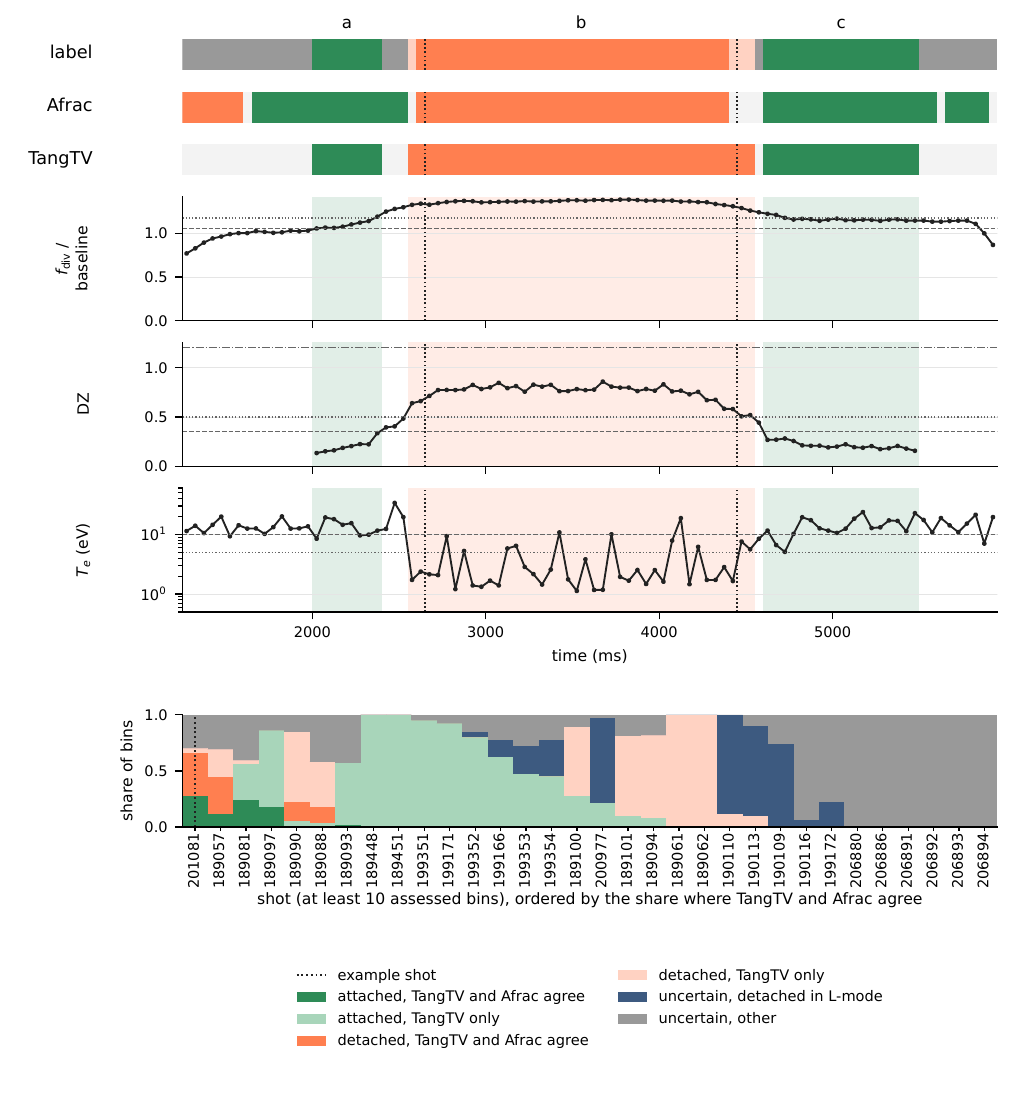}
\caption{\textbf{Detachment on shot 201081 and per shot.} Top: the label, the Afrac and
TangTV states (green attached, coral detached, gray uncertain or none), the lower
divertor's radiated fraction $f_{\mathrm{div}}$ over its baseline, front height DZ
with its cutoffs, and divertor Thomson $T_e$. a--c: attached (2000--2400\,ms),
detached (2550--4550\,ms) and attached (4600--5500\,ms) intervals; dotted: published
$T_e$ drops at 2650 and 4450\,ms. Bottom: share of each assessed shot's bins by label
and source.}
\label{fig:detach_timeline}
\end{figure*}

\section{Benchmark Models}
\label{app:models}

\paragraph{Alfv\'en eigenmodes.} \seldnet{} (ours), a SELDnet-style detector
\citep{adavanne_sound_2019}, gives per-frame activity from the four CO\textsubscript{2}
chord spectrograms (80--250\,kHz) with three 2-D convolution blocks and two
bidirectional GRUs; it is trained on 120 annotated shots and validated on the other
\auditShots{}. \lstmnet{} and \rcnnet{} \citep{garcia_comparison_2023} are an LSTM and
an echo-state network on one chord's spectrogram or a chord pair's cross-power
(20--250\,kHz), trained on the annotation; \gls{ae} activity is their largest output
other than low-frequency modes, at thresholds 0.15 and 0.10. In \cref{tab:ae_full},
``each'' rows score every chord or pair as a sample, as their evaluation does;
``averaged'' rows average a shot's inputs first.

\paragraph{Confinement.} \besnet{} rebuilds the \gls{bes} CNN of
\citet{gillRealtimePlasmaConfinement2023} from the paper, since its code and weights
were not found: log spectra of 1.02\,ms blocks from a 6$\times$8 patch of pedestal
channels, a grouped 3-D convolution and an MLP, with the paper's windows, beam
conditions, optimizer and by-shot splits. \nConfBESkept{} of \nConfBESfetched{}
fetched shots have usable \gls{bes} geometry, and the paper's test shots are unknown.
\confours{} (ours, experimental) is an untuned 1-D U-Net on D$\alpha$, line density,
$\beta_N$, stored energy and beam power, in run-day folds; its labels were set from
the traces it reads, so its score is circular.

\paragraph{Edge localized modes.} \elmours{} (ours) is a 1-D U-Net over three
D$\alpha$ filterscopes and two fast density chords, with checkpoint and threshold
picked on inner validation in five shot-grouped folds. \elmdsm{} is a detection head
on the architecture of an \gls{elm} survival model \citep{butt_wide-window_2025},
refitted on our labels with 60 of its 124 inputs. \elmo{} reimplements ELM-O
\citep{oshea_automatic_2022}, matched sample for sample to the public code: an
\gls{elm} needs an interferometer chord, two of three filterscopes and \gls{bes} to
fire, so the 46 reviewed shots without usable \gls{bes} are not scored.

\paragraph{Tearing modes.} \tmours{} (ours) is a dilated 1-D CNN on 33 Mirnov features
per 10\,ms bin (band powers, $n{=}1,2,3$ line amplitudes, and the strongest line's
frequency and $n$) with $\pm$140\,ms of centered context; it does not read the RMS the
labels come from. \tmcnn{} \citep{seoMultimodalPredictionTearing2023} and \tmdsm{}
\citep{farre-kagaInterpretingAIFusion2026} are forecasters retrained with published
hyperparameters to detect a mode at the current time, on about 324 shots per fold.
\tmrule{} scores $\max(\mathrm{RMS}_{n=1}/12\,\mathrm{G},
\mathrm{RMS}_{n=2}/6\,\mathrm{G})$.

\paragraph{Sawtooth.} \sawours{} (ours) is a PhaseNet-style picker
\citep{zhu_phasenet_2019} over 48 \gls{ece} channels and the derivative picker's two
scores, with 400\,ms of context; a second head says whether a train is present.
\sawhl{} \citep{ouyang2025realtime} is a CNN with a bidirectional LSTM, refitted on our
labels and folds; its inputs are \sawHlInputs, with the core and edge chords picked
from the data since their geometry is unknown. \sawder{} marks crashes as steep drops
in one core channel's time derivative, and a bin present within $\pm$125\,ms of one.

\paragraph{Resistive wall modes.} \rwmbrf{} and \rwmours{} are balanced random forests
\citep{piccione_predicting_2022} of 300 trees of depth 8 on a 10\,ms grid. \rwmbrf{}
reads equilibrium scalars ($\beta_N$, $l_i$, $q_{95}$, $q_{\min}$, $W_{\mathrm{MHD}}$,
$I_p$, $\beta_N/l_i$), the $n{=}1$ and $n{=}2$ magnetic RMS and toroidal rotation,
none of which senses the \gls{rwm} directly; \rwmours{} adds the $n{=}1$ $\delta B_p$ of
\cref{app:rwm}, I-coil activity and rotation at $q{=}2$. Both train only on listed
shots, and folds hold out one of three run days.

\section{Additional Results}
\label{app:results}

Brackets are 95\% shot-bootstrap intervals throughout.

\begin{table*}[ht]
\caption{\textbf{\gls{ae} benchmark in full} on the \aeBenchFrames{} frames of the
\aeBenchShots{} benchmark shots. Precision, recall and F1 at each detector's own
threshold; best F1 over all thresholds is optimistic. Each panel's last row:
\seldnet{} on all \auditShots{} validation shots.}
\label{tab:ae_full}
\centering
\small
\resizebox{\textwidth}{!}{
\begin{tabular}{lrcccccc}
\toprule
Method & Frames & AUROC & AUPRC & Precision & Recall & F1 & Best F1 \\
\midrule
\multicolumn{8}{l}{\emph{against the dense labels}} \\
\seldnet & 3{,}800 & 0.982 {\scriptsize[0.96, 0.99]} & 0.992 & 0.982 & 0.908 & 0.944 {\scriptsize[0.90, 0.98]} & 0.948 \\
\rcnnet, chord spectrogram, each chord & 15{,}200 & 0.836 {\scriptsize[0.79, 0.88]} & 0.918 & 0.893 & 0.682 & 0.774 {\scriptsize[0.73, 0.81]} & 0.849 \\
\rcnnet, chord spectrogram, 4 chords averaged & 3{,}800 & 0.862 {\scriptsize[0.81, 0.91]} & 0.930 & 0.908 & 0.699 & 0.790 {\scriptsize[0.74, 0.84]} & 0.871 \\
\rcnnet, chord-pair cross-power, each pair & 38{,}000 & 0.767 {\scriptsize[0.72, 0.82]} & 0.886 & 0.859 & 0.628 & 0.725 {\scriptsize[0.68, 0.77]} & 0.816 \\
\rcnnet, chord-pair cross-power, 10 pairs averaged & 3{,}800 & 0.778 {\scriptsize[0.73, 0.84]} & 0.896 & 0.857 & 0.660 & 0.746 {\scriptsize[0.69, 0.80]} & 0.827 \\
\lstmnet, chord spectrogram, each chord & 15{,}200 & 0.714 {\scriptsize[0.65, 0.77]} & 0.842 & 0.785 & 0.656 & 0.715 {\scriptsize[0.68, 0.75]} & 0.838 \\
\lstmnet, chord spectrogram, 4 chords averaged & 3{,}800 & 0.736 {\scriptsize[0.66, 0.80]} & 0.855 & 0.798 & 0.686 & 0.738 {\scriptsize[0.69, 0.79]} & 0.851 \\
\lstmnet, chord-pair cross-power, each pair & 38{,}000 & 0.699 {\scriptsize[0.61, 0.77]} & 0.825 & 0.793 & 0.626 & 0.700 {\scriptsize[0.65, 0.74]} & 0.829 \\
\lstmnet, chord-pair cross-power, 10 pairs averaged & 3{,}800 & 0.714 {\scriptsize[0.61, 0.80]} & 0.841 & 0.772 & 0.629 & 0.693 {\scriptsize[0.65, 0.74]} & 0.841 \\
always present & 3{,}800 & 0.500 & 0.683 & 0.683 & 1.000 & 0.812 & 0.812 \\
\seldnet{} on all 60 validation shots (reference) & 12{,}000 & 0.980 & 0.992 & 0.974 & 0.928 & 0.951 & 0.954 \\
\midrule
\multicolumn{8}{l}{\emph{against the annotation}} \\
\seldnet & 3{,}800 & 0.680 {\scriptsize[0.54, 0.80]} & 0.351 & 0.297 & 0.894 & 0.445 {\scriptsize[0.34, 0.57]} & 0.448 \\
\rcnnet, chord spectrogram, each chord & 15{,}200 & 0.896 {\scriptsize[0.86, 0.93]} & 0.667 & 0.391 & 0.975 & 0.559 {\scriptsize[0.46, 0.65]} & 0.651 \\
\rcnnet, chord spectrogram, 4 chords averaged & 3{,}800 & 0.911 {\scriptsize[0.87, 0.94]} & 0.691 & 0.399 & 1.000 & 0.570 {\scriptsize[0.47, 0.66]} & 0.676 \\
\rcnnet, chord-pair cross-power, each pair & 38{,}000 & 0.867 {\scriptsize[0.82, 0.91]} & 0.585 & 0.393 & 0.935 & 0.553 {\scriptsize[0.45, 0.64]} & 0.614 \\
\rcnnet, chord-pair cross-power, 10 pairs averaged & 3{,}800 & 0.886 {\scriptsize[0.83, 0.93]} & 0.616 & 0.388 & 0.974 & 0.554 {\scriptsize[0.45, 0.65]} & 0.645 \\
\lstmnet, chord spectrogram, each chord & 15{,}200 & 0.857 {\scriptsize[0.81, 0.89]} & 0.598 & 0.347 & 0.946 & 0.508 {\scriptsize[0.43, 0.58]} & 0.599 \\
\lstmnet, chord spectrogram, 4 chords averaged & 3{,}800 & 0.898 {\scriptsize[0.85, 0.93]} & 0.668 & 0.354 & 0.992 & 0.522 {\scriptsize[0.44, 0.59]} & 0.647 \\
\lstmnet, chord-pair cross-power, each pair & 38{,}000 & 0.867 {\scriptsize[0.83, 0.89]} & 0.558 & 0.376 & 0.968 & 0.541 {\scriptsize[0.46, 0.62]} & 0.608 \\
\lstmnet, chord-pair cross-power, 10 pairs averaged & 3{,}800 & 0.892 {\scriptsize[0.86, 0.92]} & 0.607 & 0.374 & 0.995 & 0.544 {\scriptsize[0.45, 0.62]} & 0.654 \\
always present & 3{,}800 & 0.500 & 0.209 & 0.209 & 1.000 & 0.346 & 0.346 \\
\seldnet{} on all 60 validation shots (reference) & 12{,}000 & 0.625 & 0.275 & 0.260 & 0.882 & 0.402 & 0.407 \\
\bottomrule
\end{tabular}
}
\end{table*}

\begin{figure*}[ht]
\centering
\includegraphics[width=\textwidth]{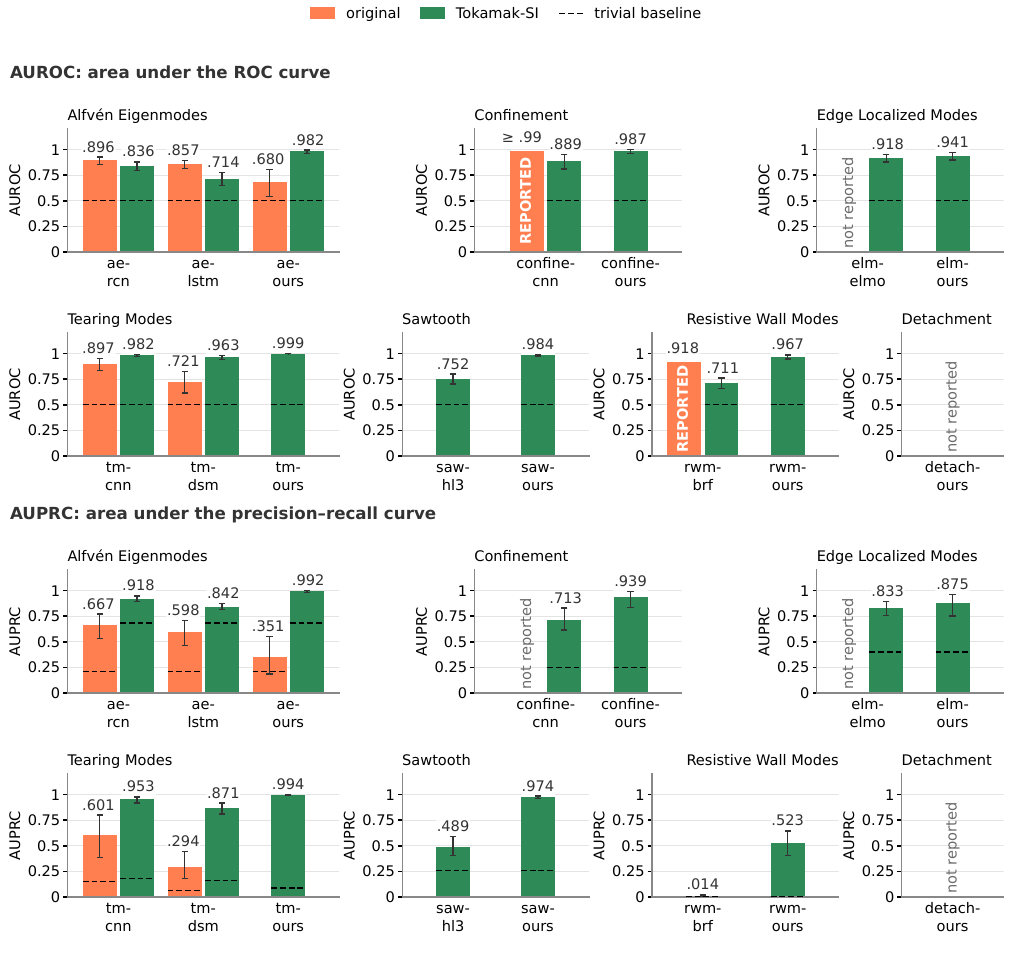}
\caption{\textbf{AUROC (rows 1--2) and AUPRC (rows 3--4)} in the settings of
\cref{fig:benchmarks}. The confinement REPORTED bar is its paper's bound of at least
0.99 per class; the \gls{rwm} REPORTED bar is the NSTX forest's forecast AUROC on NSTX.
Dashes: trivial baselines.}
\label{fig:benchmarks_auc}
\end{figure*}

\begin{table*}[ht]
\caption{\textbf{Confinement benchmark in full.} Top: \besnet{} under its paper's
protocol, pooled over five random by-shot splits (one split scores
\besOneSplit$\pm$\besOneSplitSD); last column: its paper's scores. \gls{wpqh} rests on
\nConfWPtest{} test shots. Bottom: \confours{}, experimental and circular. On
\besnet's test windows, \confours{} minus \besnet{} is \confOursPaired{}
\confOursPairedci.}
\label{tab:conf_bes}
\centering
\small
\begin{tabular}{lrcccccc}
\toprule
Regime & Windows (test shots) & Precision & Recall & F1 & AUROC & AUPRC & Paper F1 (P / R) \\
\midrule
\multicolumn{8}{l}{\besnet{}: five random by-shot splits under the paper's protocol, pooled (\nConfBESshots{} distinct shots, \nConfBEStests{} shot-tests)} \\
L-mode & 23{,}693 (40) & 0.847 & 0.750 & 0.795 {\scriptsize[0.64, 0.89]} & 0.937 & 0.808 & 0.94 (0.98 / 0.90) \\
H-mode & 79{,}557 (69) & 0.841 & 0.875 & 0.858 {\scriptsize[0.78, 0.91]} & 0.925 & 0.940 & 0.97 (0.95 / 0.98) \\
\gls{qh}-mode & 44{,}061 (59) & 0.744 & 0.790 & 0.766 {\scriptsize[0.66, 0.85]} & 0.912 & 0.781 & 0.94 (0.92 / 0.96) \\
\gls{wpqh}-mode & 5{,}552 (7) & 0.658 & 0.278 & \besWPFone{} {\scriptsize\besWPci} & 0.785 & 0.324 & 0.90 (0.95 / 0.86) \\
Macro & 152{,}863 (\nConfBESshots) & 0.773 & 0.673 & \besMacroFone{} {\scriptsize\besMacroCI} & \besMacroAUROC{} {\scriptsize\besMacroAUROCci} & \besMacroAUPRC & 0.94 (0.95 / 0.93) \\
\midrule
\multicolumn{8}{l}{\confours{}: 1\,ms bins, 5-fold cross-validation holding out experiment days (\nConfOursShots{} shots)} \\
L-mode & 113{,}862 (113) & & & 0.924 {\scriptsize[0.87, 0.96]} & 0.965 & 0.942 & \\
H-mode & 408{,}762 (166) & & & 0.968 {\scriptsize[0.95, 0.98]} & 0.994 & 0.991 & \\
\gls{qh}-mode & 221{,}977 (186) & & & 0.914 {\scriptsize[0.88, 0.94]} & 0.984 & 0.951 & \\
\gls{wpqh}-mode & 95{,}416 (71) & & & 0.836 {\scriptsize[0.76, 0.90]} & 0.976 & 0.907 & \\
Macro & 840{,}017 (\nConfOursShots) & & & \confOursFone{} {\scriptsize\confOursCI} & \confOursAUROC & \confOursAUPRC & \\
\bottomrule
\end{tabular}

\end{table*}

\begin{table}[ht]
\caption{\textbf{\gls{elm} benchmark}, cross-validated on 50\,ms bins wholly inside one
reviewed span. On the \gls{bes} shots, \elmours{} minus \elmo{} is AUROC \elmPairA{}
\elmPairAci{} and F1 \elmPairF{} \elmPairFci. Alarm: share of absent spans a detection
touches, also with 25\,ms trimmed from each edge. The peak finder seeded the review;
the last block keeps spans whose edges the experts moved by more than 1\,ms.
$^\dagger$Recall $\geq$0.99.}
\label{tab:elm_benchmark}
\centering
\scriptsize\setlength{\tabcolsep}{2.5pt}
\begin{tabular}{lcccccc}
\toprule
Method & AUROC & AUPRC & F1 & Absent-bin FPR & Absent-span alarm & Alarm, trimmed \\
\midrule
\multicolumn{7}{l}{All reviewed: 119 shots/12,409 bins} \\
\texttt{elm-ours} & 0.941 {\scriptsize[0.906, 0.967]} & 0.877 {\scriptsize[0.774, 0.949]} & 0.830 {\scriptsize[0.778, 0.871]} & 0.113 {\scriptsize[0.075, 0.152]} & 0.596 {\scriptsize[0.478, 0.689]} & 0.427 {\scriptsize[0.343, 0.495]} \\
\texttt{elm-peaks} & -- & -- & 0.757 {\scriptsize[0.680, 0.827]} & 0.063 {\scriptsize[0.032, 0.100]} & 0.105 {\scriptsize[0.063, 0.168]} & 0.096 {\scriptsize[0.056, 0.154]} \\
\texttt{always-present} & 0.500 {\scriptsize[0.500, 0.500]} & 0.376 {\scriptsize[0.322, 0.430]} & 0.546 {\scriptsize[0.487, 0.601]}$^{\dagger}$ & 1.000 {\scriptsize[1.000, 1.000]} & 1.000 {\scriptsize[1.000, 1.000]} & 0.915 {\scriptsize[0.845, 0.976]} \\
\midrule
\multicolumn{7}{l}{BES subset: 73 shots/6,843 bins} \\
\texttt{elm-ours} & 0.941 {\scriptsize[0.898, 0.972]} & 0.875 {\scriptsize[0.751, 0.961]} & 0.845 {\scriptsize[0.786, 0.896]} & 0.123 {\scriptsize[0.071, 0.184]} & 0.610 {\scriptsize[0.452, 0.742]} & 0.436 {\scriptsize[0.351, 0.511]} \\
\texttt{elm-elmo} & 0.918 {\scriptsize[0.877, 0.951]} & 0.833 {\scriptsize[0.753, 0.890]} & 0.842 {\scriptsize[0.789, 0.885]} & 0.108 {\scriptsize[0.068, 0.156]} & 0.440 {\scriptsize[0.301, 0.623]} & 0.307 {\scriptsize[0.200, 0.444]} \\
\texttt{elm-peaks} & -- & -- & 0.708 {\scriptsize[0.602, 0.800]} & 0.091 {\scriptsize[0.039, 0.146]} & 0.110 {\scriptsize[0.053, 0.188]} & 0.106 {\scriptsize[0.051, 0.182]} \\
\texttt{always-present} & 0.500 {\scriptsize[0.500, 0.500]} & 0.402 {\scriptsize[0.325, 0.479]} & 0.573 {\scriptsize[0.491, 0.648]}$^{\dagger}$ & 1.000 {\scriptsize[1.000, 1.000]} & 1.000 {\scriptsize[1.000, 1.000]} & 0.890 {\scriptsize[0.788, 0.982]} \\
\midrule
\multicolumn{7}{l}{BES subset, edges moved $>1$ ms from the peak finder's: 43 present and 117 absent spans/4,125 bins} \\
\texttt{elm-ours} & 0.938 {\scriptsize[0.875, 0.980]} & 0.836 {\scriptsize[0.629, 0.972]} & 0.837 {\scriptsize[0.738, 0.910]} & 0.096 {\scriptsize[0.043, 0.168]} & -- & -- \\
\texttt{elm-elmo} & 0.934 {\scriptsize[0.894, 0.965]} & 0.816 {\scriptsize[0.688, 0.909]} & 0.822 {\scriptsize[0.730, 0.884]} & 0.098 {\scriptsize[0.048, 0.163]} & -- & -- \\
\texttt{elm-peaks} & -- & -- & 0.516 {\scriptsize[0.297, 0.682]} & 0.136 {\scriptsize[0.061, 0.225]} & -- & -- \\
\bottomrule
\end{tabular}

\end{table}

\begin{table*}[ht]
\caption{\textbf{Tearing-mode benchmark}: recovery of the magnetic rule. F1 uses
inner-validation thresholds except the CNN rows marked thr.\ 0.5 (its card's); the
saved DSM has no published threshold. Segmental F1 matches segments at IoU 0.5. Top:
original targets. Middle and bottom: uncertain time excluded or scored as negative.
``Published'' rows are the saved forecasters, ``retrained'' rows the same
architectures retrained for detection. $\dagger$Training overlap unknown.}
\label{tab:tm_benchmark}
\centering
\scriptsize
\setlength{\tabcolsep}{3pt}
\begin{tabular}{>{\raggedright\arraybackslash}p{1.75in}rllll}
\toprule
Model & Shots / bins & AUROC [CI] & AUPRC [CI] & F1 [CI] & Segmental F1 [CI] \\
\midrule
\multicolumn{6}{l}{\emph{Original targets: CNN at its card's threshold, DSM at its inner-validation threshold}} \\
\texttt{tm-cnn-published} (thr. 0.5)$\dagger$ & 79 / 5.6k & .897 [.832,.952] & .601 [.385,.799] & .611 [.437,.750] & -- \\
\texttt{tm-dsm-500ms-published} & 80 / 8.4k & .721 [.614,.828] & .294 [.179,.442] & .177 [.107,.266] & -- \\
\midrule
\multicolumn{6}{l}{\emph{Tokamak-SI labels, uncertain time excluded}} \\
\texttt{tm-cnn-published}$\dagger$ & 278 / 31.0k & .974 [.959,.987] & .903 [.838,.948] & .779 [.689,.848] & .615 [.504,.713] \\
\texttt{tm-cnn-published} (thr. 0.5)$\dagger$ & 278 / 31.0k & .974 [.959,.987] & .903 [.838,.948] & .708 [.582,.808] & .588 [.416,.746] \\
\texttt{tm-dsm-500ms-published} & 247 / 43.7k & .779 [.691,.852] & .405 [.285,.528] & .442 [.327,.534] & .279 [.157,.403] \\
\texttt{tm-cnn-retrained} & 278 / 31.2k & .982 [.968,.993] & .953 [.916,.976] & .844 [.758,.906] & .683 [.538,.785] \\
\texttt{tm-dsm-retrained} & 336 / 58.1k & .963 [.940,.981] & .871 [.810,.916] & .799 [.728,.844] & .657 [.545,.745] \\
\tmrule{} & 450 / 162.8k & .998 [.998,.999] & .986 [.980,.990] & .940 [.926,.953] & .629 [.538,.717] \\
\texttt{tm-ours} & 450 / 162.8k & .999 [.997,1.000] & .994 [.989,.997] & .960 [.930,.980] & .843 [.778,.901] \\
\midrule
\multicolumn{6}{l}{\emph{Tokamak-SI labels, uncertain time scored as negative}} \\
\texttt{tm-cnn-published}$\dagger$ & 286 / 59.1k & .900 [.864,.932] & .412 [.281,.571] & .406 [.306,.497] & .288 [.191,.374] \\
\texttt{tm-cnn-published} (thr. 0.5)$\dagger$ & 286 / 59.1k & .900 [.864,.932] & .412 [.281,.571] & .516 [.396,.625] & .394 [.240,.535] \\
\texttt{tm-dsm-500ms-published} & 252 / 81.9k & .712 [.622,.795] & .157 [.098,.228] & .252 [.166,.330] & .167 [.071,.271] \\
\texttt{tm-cnn-retrained} & 286 / 59.3k & .903 [.873,.932] & .444 [.292,.599] & .435 [.331,.528] & .313 [.202,.417] \\
\texttt{tm-dsm-retrained} & 341 / 112.2k & .839 [.791,.880] & .297 [.204,.411] & .356 [.278,.427] & .262 [.175,.347] \\
\tmrule{} & 450 / 246.8k & .963 [.954,.971] & .632 [.559,.700] & .447 [.372,.519] & .191 [.152,.235] \\
\texttt{tm-ours} & 450 / 246.8k & .948 [.935,.960] & .512 [.394,.623] & .363 [.290,.431] & .229 [.182,.281] \\
\bottomrule
\end{tabular}

\end{table*}

\section{Language Interface Example}
\label{app:llm}

A \emph{get events} call on shot 189042 returns 302 events from seven sources, one
text mention and no forecasts. A caveat explains the mention:
\begin{quote}\small\ttfamily
1 row(s) are in `text\_mentions', not in `events': a text row is a LEXICON HIT in
the operator logbook -- somebody wrote a word -- and is not an assertion that the
phenomenon occurred \dots
\end{quote}
A \emph{phenomenon locate} query for ``tearing'' ranks hits by evidence class, with
logbook caveats such as ``operator log says NOT Tearing mode''. Each event records
its source and coverage, so a label source can be replaced as a unit.

\section{Datasheet Summary}
\label{app:datasheet}

\textbf{Motivation.} Dense, coverage-aware labels and a benchmark for shot
interpretation. \textbf{Composition.} Four-state interval labels for \nPhen{}
phenomena in the tiers of \cref{sec:design}, on \nPop{} rule-selected \gls{d3d} shots
and a \nCohort-shot weighted cohort, plus literature links and model suggestions.
\textbf{Collection.} Expert review on fixed panels (blind for the test split),
credited original tables, and detector suggestions. \textbf{Uses.} Training and
evaluating event detectors and interpreters; auditing references. Suggestions are
not ground truth. \textbf{Distribution.} Labels with provenance, shot lists,
selection rules, scoring code, weights and the scripts behind every number; raw
signals stay under the \gls{d3d} data policy. \textbf{Maintenance.} Numbered releases,
each stating what changed and re-scoring every model once on gold.

\clearpage
\section{Examples}
\label{app:examples}

\begin{figure*}[ht]
\centering
\includegraphics[width=\textwidth]{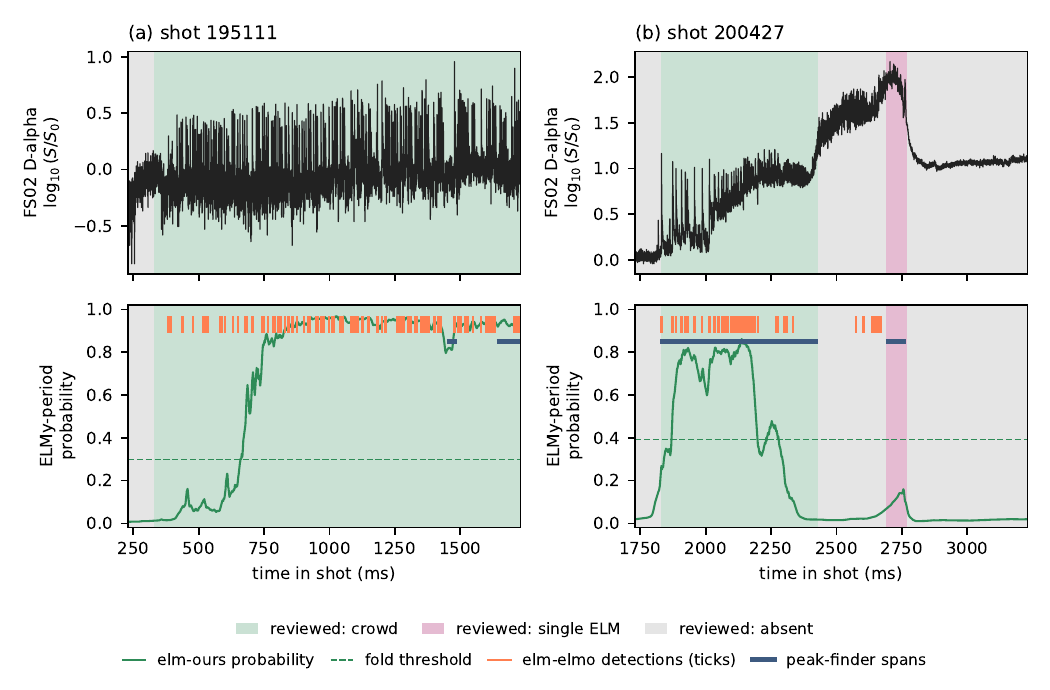}
\caption{\textbf{\gls{elm} labels and detectors} on shots 195111 (a) and 200427 (b), at
the 75th and 25th percentiles of \elmours{} F1. Top: D$\alpha$. Shading: reviewed
crowd, single-\gls{elm} and absent spans. Bottom: \elmours{} probability and
threshold, \elmo{} detections and the peak finder's spans. In (b), the single-\gls{elm}
span at 2689--2765\,ms lies in a high-recycling phase, so \gls{elm} identity there is
ambiguous.}
\label{fig:elm_examples}
\end{figure*}

\begin{figure}[ht]
\centering
\includegraphics[width=0.48\textwidth]{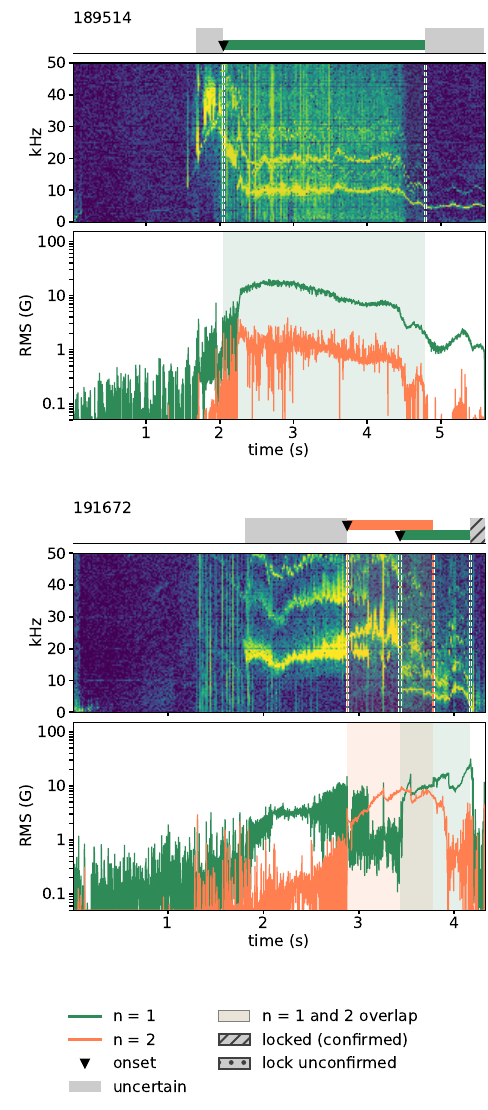}
\caption{\textbf{Tearing-mode labels} on shots 189514 (a mode that decays) and 191672
(an $n{=}2$ mode, then an $n{=}1$ mode that locks): the label bar (onsets as
triangles, uncertain time gray, a confirmed lock hatched), a Mirnov spectrogram, and
the $n{=}1$ and $n{=}2$ magnetic RMS.}
\label{fig:tm_examples}
\end{figure}

\end{document}